# Logical Foundations and Complexity of 4QL, a Query Language with Unrestricted Negation

**Jan Małuszyński[a] and Andrzej Szałas[a,b]**

[a] *Department of Computer and Information Science, University of Linköping, SE-581 83 Linköping, Sweden, email: jan.maluszynski@ida.liu.se.*
[b] *Institute of Informatics, Warsaw University, Banacha 2, 02-097 Warsaw, Poland email: andrzej.szalas@mimuw.edu.pl*

ABSTRACT. *The paper discusses properties of a* DATALOG$^{\neg\neg}$*-like query language 4QL, originally outlined by Małuszyński and Szałas [MS11]. 4QL allows one to use rules with negation in heads and bodies of rules. It is based on a simple and intuitive semantics and provides uniform tools for "lightweight" versions of known forms of nonmonotonic reasoning. Negated literals in heads of rules may naturally lead to inconsistencies. On the other hand, rules do not have to attach meaning to some literals. Therefore 4QL is founded on a four-valued semantics, employing the logic introduced in [MSV08, VMS09] with truth values: 'true', 'false', 'inconsistent' and 'unknown'. In addition, 4QL is tractable w.r.t. data complexity and captures* PTIME *queries. Even though* DATALOG$^{\neg\neg}$ *is known as a concept for the last 30 years, to our best knowledge no existing approach enjoys these properties.*
*In the current paper we:*

- *investigate properties of well-supported models of 4QL*
- *prove the correctness of the algorithm for computing well-supported models*
- *show that 4QL has* PTIME *data complexity and captures* PTIME.

## 1. Introduction and Preliminaries

This is a companion paper to our conference paper [MS11]. It is motivated by the problem of handling explicit negative knowledge in a rule language. This means that negation can appear in bodies and heads of rules, as done in DATALOG$^{\neg\neg}$ (see, e.g., [AHV96]). However, in contrast to the traditional query languages based on nonmonotonic logics initially derived from the Closed World Assumption [Rei78, AD98] (CWA) we focus our attention on explicit negative knowledge. In [MS11] we outlined a novel lightweight approach to this problem, separating the issues of incomplete/inconsistent knowledge and nonmonotonicity. To make separation between monotonic-nonmonotonic clear, we first follow the Open World Assumption (OWA) in the presence of the explicit negation and later provide simple con-structs allowing one to structure knowledge bases in the form of modules and to query such modules. As shown in [MS11], these constructs provide a natural way to express CWA and “lightweight” versions of other known forms of nonmonotonic reasoning.

These methodological assumptions led to a DATALOG$^{\neg\neg}$-like query language, 4QL, founded on a four-valued semantics. The semantics employs the logic introduced in [MSV08, VMS09] with truth values: ‘true’, ‘false’, ‘inconsistent’ and ‘unknown’, further denoted by $\mathfrak{t}$, $\mathfrak{f}$, $\mathfrak{i}$ and $\mathfrak{u}$, respectively.

For the four-valued query language 4QL proposed in [MS11] we defined a semantics based on the notion of well-supported models and we proposed an algorithm for computing such models. The language:

- allows one to use rules with unrestricted negation (in heads and bodies)
- has a simple and intuitive semantics
- allows for “lightweight” versions of known forms of nonmonotonic reasoning.

Due to page limits, [MS11] does not contain important properties of well-supported models, nor proofs of theorems and complexity analysis. In the current paper we therefore:

- investigate properties of well-supported models of 4QL
- prove the correctness of the algorithm for computing well-supported models
- show that 4QL has PTIME data complexity and captures PTIME.

Even though DATALOG$^{\neg\neg}$ is known as a concept for 30 years, to our best knowledge no existing approach enjoys these properties.

In Section 2 we discuss related work. Next, in Section 3, we provide syntax and semantics of 4QL via the notion of well-supported models and continue in Section 4 with properties of well-supported models. Section 5 is devoted to an algorithm for computing well-supported models and a proof of its correctness. In Section 6 we extend basic 4QL by very simple constructs which allow us to show in Section 7 that the extended 4QL captures PTIME on ordered databases. Finally, Section 8 concludes the paper.

## 2. Related Work

While DATALOG$^{\neg}$ has attracted much attention (see, e.g., [AHV96] and references therein), a tractable and at the same time intuitive semantics for DATALOG$^{\neg\neg}$ is not reported in the literature. The problem is often addressed by taking DATALOG with negation or extended logic programs [GL91] as the starting point for paraconsistent extensions, see e.g. [ADP05, dAP07, Ari02, DP98, SI95].[1]

Let us now discuss approaches [ADP05, Ari02, dAP07, Fit02] which are closest to ours.

The paper [ADP05] provides a framework for paraconsistent logic programming generalizing some previous work of its authors to an arbitrary complete bilattice of truth-values, where belief and doubt are explicitly represented. In contrast to our approach, where default

1. There is a rich literature on paraconsistent logics (see, e.g., [BCG07] and references therein, in particular [CMM07]).

negation is not used, the focus of this work is semantical integration of explicit and default negation.

The paper [Ari02] proposes a framework for dealing with logic programs with two kinds of negation: an explicit negation and a negation-by-failure. We use a single negation, without referring to provability, as the negation-as-failure does. We find this an advantage of our approach. Also our implication is different. As [Ari02] concentrates on logic programs, database related topics are not considered there. In particular no complexity results concerning finite domains are provided.

Another approach, P-Datalog [dAP07], provides a paraconsistent language for knowledge base integration based on a four-valued logic with the truth ordering which coincides with our truth ordering. However, there are several important differences. First, our language only allows explicit negation (in the head and in the bodies of the rules) while P-Datalog programs only allow negation by default $\sim$ in the rule bodies and use the CWA. Information integration done in P-Datalog uses the truth ordering, while we provide a more general constructs for achieving this goal (extended literals, defined in Section 6). Finally, the rules are interpreted by substantially different implications.

A four-valued semantics for logic programs with negation was also proposed by Fitting (see e.g. [Fit02]). The major difference is that Fitting's semantics is based on the Belnap's logic. This contrasts with our approach since we use a different truth ordering. Second, the semantics of Fitting allows one to derive conclusions from false premises, which we do not allow. Unlike in our framework, a rule of a Fitting program is satisfied if and only if truth values assigned to the head and to the body are equal. Last and not least, the language considered by Fitting does not admit negation in rule heads.

A major difficulty in defining 4QL was handling of disjunction in rule bodies. This paper presents a solution to this problem. A preliminary idea of a four-valued rule language not admitting disjunction in rule bodies appeared in our previous work [VMS09].

Also in the field of the Semantic Web there have been proposals addressing the problem of inconsistent data by defining paraconsistent extensions of Description Logics underlying the Semantic Web Ontology Language OWL2, see, e.g., [MS96, Str97, OW08, MH09, Mai10] (where the issue of nonmonotonicity is not addressed) or by providing rule languages based on defeasible reasoning [Nut94].

## 3. Syntax and semantics of 4QL

### 3.1. *Syntax of 4QL*

For a negative literal $\ell$ the notation $\neg\ell$ denotes its positive counterpart. We treat propositions as zero-argument relation symbols.

In the sequel, for simplicity, we consider ground rules only and assume that for each head $\ell$ there is only one rule having the form:

$$\ell :- (b_{11}, \ldots, b_{1i_1}) \vee (b_{21}, \ldots, b_{2i_2}) \vee \ldots \vee (b_{m1}, \ldots, b_{mi_m}). \quad (1)$$

The rationale for this syntax is as follows. In two-valued logic conjunction of the classical implications of the form $a \rightarrow \ell$ and $b \rightarrow \ell$ is logically equivalent to the implication $a \vee b \rightarrow \ell$. Thus, multiple DATALOG rules with the same rule heads express the disjunction of bodies of these rules. In logic programming this is sometimes used to avoid explicit disjunction in rule bodies. In our four-valued logic the equivalence does not hold (see Remark 4). Therefore, in order to have disjunction in the rule body we have to express it explicitly therein.

In 4QL one can consider admission of multiple rules with the same rule heads, interpreted as four-valued conjunction of such rules. However, intuitive understanding of such four-valued rules is rather difficult. Intuitively we consider a rule with head $\ell$ as a separate information source concerning $\ell$. We prefer to leave to the user how information from different sources is to be combined rather than to impose four-valued conjunction as the only way of combination. Therefore at the basic level of 4QL we do not admit multiple rules with the same head. In the sequel we extend 4QL with a flexible mechanism for fusing information from different sources.

The empty body of a rule is denoted by $\emptyset$. We sometimes refer to *facts* as to rules of the form $\ell$ :– $\emptyset$.

DEFINITION 1. — *Let $\varrho$ be a rule of the form* (1)*. Then:*

– $rule(\ell) \stackrel{\text{def}}{=} \varrho$
– $head(\varrho) \stackrel{\text{def}}{=} \ell$
– $body(\varrho) \stackrel{\text{def}}{=} (b_{11}, \ldots, b_{1i_1}) \vee (b_{21}, \ldots, b_{2i_2}) \vee \ldots \vee (b_{m1}, \ldots, b_{mi_m})$
– *for* $1 \leq j \leq m$, $\beta_j(\varrho) \stackrel{\text{def}}{=} b_{j1}, \ldots, b_{ji_j}$.

Writing $rule(l)$, we always assume that the respective rule with $\ell$ as its head indeed exists.

REMARK 2. — In the paper we present the case of ground rules only. However, typical rules with variables are allowed, too. We assume that whenever there is a variable appearing in the body of a rule but not in its head then it is assumed to be existentially quantified in its body. For example,

$$p(x, y) :\!-\, q(x, y, z).$$

is understood as "$p(x, y) :\!-\, \exists z[q(x, y, z)]$". The existential quantifier is then understood as the disjunction $q(x, y, a_1) \vee \ldots \vee q(x, y, a_k)$, where $a_1, \ldots, a_k$ are all constants appearing in the database. □

## 3.2. *The Underlying Logic*

Addressing the semantics of DATALOG$^{\neg\neg}$ we have made certain methodological choices. First of all, we decided to start with a fully monotonic query language. Therefore, rather than starting with CWA, as most approaches do, we have accepted OWA. This, in turn, naturally introduces $\mathfrak{u}$ as a truth value. For example, having just one rule $p$ :– $q$ we do not force $p$ and $q$ to be $\mathfrak{f}$, so have to assign them the value $\mathfrak{u}$. Next, since we allow negated literals in heads of rules, certain conclusions may contradict each other. This also gives rise to a non-classical truth value. Could this value be $\mathfrak{u}$? If one does not want to distinguish between the lack of information and inconsistency, such an identification could do the job (see, e.g., [dACM02, CMdA00, DŁSS06, DMS06, Fit02, Fit85, NS10]). However, we find it more natural and informative to distinguish between $\mathfrak{u}$ and $\mathfrak{i}$. We then adopt the four-valued logic introduced in [MSV08] as the semantical foundation. The truth tables for the connectives are shown in Table 1.

**Table 1.** *Truth tables for* $\wedge$, $\vee$, $\rightarrow$ *and* $\neg$.

| $\wedge$ | $\mathfrak{f}$ | $\mathfrak{u}$ | $\mathfrak{i}$ | $\mathfrak{t}$ |
|---|---|---|---|---|
| $\mathfrak{f}$ | $\mathfrak{f}$ | $\mathfrak{f}$ | $\mathfrak{f}$ | $\mathfrak{f}$ |
| $\mathfrak{u}$ | $\mathfrak{f}$ | $\mathfrak{u}$ | $\mathfrak{u}$ | $\mathfrak{u}$ |
| $\mathfrak{i}$ | $\mathfrak{f}$ | $\mathfrak{u}$ | $\mathfrak{i}$ | $\mathfrak{i}$ |
| $\mathfrak{t}$ | $\mathfrak{f}$ | $\mathfrak{u}$ | $\mathfrak{i}$ | $\mathfrak{t}$ |

| $\vee$ | $\mathfrak{f}$ | $\mathfrak{u}$ | $\mathfrak{i}$ | $\mathfrak{t}$ |
|---|---|---|---|---|
| $\mathfrak{f}$ | $\mathfrak{f}$ | $\mathfrak{u}$ | $\mathfrak{i}$ | $\mathfrak{t}$ |
| $\mathfrak{u}$ | $\mathfrak{u}$ | $\mathfrak{u}$ | $\mathfrak{i}$ | $\mathfrak{t}$ |
| $\mathfrak{i}$ | $\mathfrak{i}$ | $\mathfrak{i}$ | $\mathfrak{i}$ | $\mathfrak{t}$ |
| $\mathfrak{t}$ | $\mathfrak{t}$ | $\mathfrak{t}$ | $\mathfrak{t}$ | $\mathfrak{t}$ |

| $\rightarrow$ | $\mathfrak{f}$ | $\mathfrak{u}$ | $\mathfrak{i}$ | $\mathfrak{t}$ |
|---|---|---|---|---|
| $\mathfrak{f}$ | $\mathfrak{t}$ | $\mathfrak{t}$ | $\mathfrak{t}$ | $\mathfrak{t}$ |
| $\mathfrak{u}$ | $\mathfrak{t}$ | $\mathfrak{t}$ | $\mathfrak{t}$ | $\mathfrak{t}$ |
| $\mathfrak{i}$ | $\mathfrak{f}$ | $\mathfrak{f}$ | $\mathfrak{t}$ | $\mathfrak{f}$ |
| $\mathfrak{t}$ | $\mathfrak{f}$ | $\mathfrak{f}$ | $\mathfrak{t}$ | $\mathfrak{t}$ |

| | $\neg$ |
|---|---|
| $\mathfrak{f}$ | $\mathfrak{t}$ |
| $\mathfrak{u}$ | $\mathfrak{u}$ |
| $\mathfrak{i}$ | $\mathfrak{i}$ |
| $\mathfrak{t}$ | $\mathfrak{f}$ |

Notice that our logic is different from the commonly used four-valued Belnap's logic [Bel77]. In Belnap's logic two orderings on truth values are considered, known as *knowledge ordering* and *truth ordering*. As shown, e.g., in [Dub08] as well as in our previous work [VMS09, NS10], Belnap's truth ordering is problematic in areas we focus on and we introduced instead a different truth ordering $\mathfrak{f} < \mathfrak{u} < \mathfrak{i} < \mathfrak{t}$, which was also independently proposed in [dAP07]. The truth tables for conjunction $\wedge$ and disjunction $\vee$ are respectively defined as minimum and maximum w.r.t. our truth ordering.

REMARK 3. — To motivate our truth ordering, where in particular $\mathfrak{u} < \mathfrak{i}$, first note that $\mathfrak{i}$ contains an evidence that a given proposition is $\mathfrak{t}$ so in this respect it is "closer" to truth than $\mathfrak{u}$. On the other hand, one can argue that $\mathfrak{i}$ reflects some evidence that the corresponding assertion is not true, while $\mathfrak{u}$ does not. Observe however, that in the monotonic layer we do not want to derive conclusions from unknown premises, since such derivations lead to nonmonotonicity. In contrary, we want to derive conclusions from inconsistent premises to "correct" heads of rules. To illustrate this point consider the following rule:

$$reduce_temperature :\!- \; high_temperature.$$

If, in a given interpretation, $high_temperature$ is $\mathfrak{u}$ then one does not want to be forced to derive any conclusions whether to reduce temperature or not, unless nonmonotonic reasoning is used (which is addressed in Section 6).

Consider now a situation, when at some point $high_temperature$ is $\mathfrak{t}$. Then we also want to conclude that $reduce_temperature$ is $\mathfrak{t}$. If the situation changes and $high_temperature$ becomes $\mathfrak{i}$, e.g., by deriving new facts then the previous conclusion as to the truth of $reduce_temperature$ is no longer justified and is to be adjusted to $\mathfrak{i}$.

The above case indicates that $\mathfrak{i}$ behaves like $\mathfrak{t}$ (so can be considered as a designated value), while $\mathfrak{u}$ does not. This motivates our choice as to $\mathfrak{u} < \mathfrak{i}$. □

The implication $\rightarrow$ is a four-valued extension of the classical implication. While its arguments range over the four logical values the truth value of the implication is $\mathfrak{t}$ or $\mathfrak{f}$. The implication is used to interpret clauses of 4QL. Whenever the body of a clause has the value $\mathfrak{f}$ or $\mathfrak{u}$, the truth value of the clause is defined to be $\mathfrak{t}$. Intuitively, this reflects our intention not to draw conclusions from false or unknown information. Namely, a clause with unknown or false body is always satisfied, so one does not have to update its head. From inconsistent body we want to conclude that the head is also inconsistent. Thus, for the predecessor with value $\mathfrak{i}$ the implication is $\mathfrak{t}$ if the successor is $\mathfrak{i}$, and $\mathfrak{f}$ otherwise. The implication is $\mathfrak{t}$ if the predecessor takes value $\mathfrak{t}$ and the successor is $\mathfrak{t}$ or $\mathfrak{i}$. The latter case is needed to handle the situation when both head and its negation are to be derived on the basis of true assumptions.

REMARK 4. — Note that the classical equivalence

$$[(p \to q) \wedge (r \to q)] \equiv [(p \vee r) \to q],$$

allowing one to consider several rules as a single one, does not hold in the four-valued setting we deal with. For example, $(\mathfrak{i} \to \mathfrak{t}) \wedge (\mathfrak{t} \to \mathfrak{t})$ is $\mathfrak{t}$, while $(\mathfrak{i} \vee \mathfrak{t}) \to \mathfrak{t}$ is $\mathfrak{f}$.

Observe also that that the negation defined in Table 1 is not truth reversing ($\mathfrak{i} > \mathfrak{u}$ but not $\neg\,\mathfrak{i} < \neg\mathfrak{u}$). Our definition of $\neg$ reflects the intuition that when we have no information as to the truth or falsity of $p$ then we still have no such information about $\neg p$. This usually is not questioned in the three-valued case without $\mathfrak{i}$, where $\neg\mathfrak{u} \stackrel{\text{def}}{=} \mathfrak{u}$. Similarly, having inconsistent $p$, we still have inconsistent $\neg p$. In order to have a truth reversing negation, we would have to accept that $\neg\mathfrak{i} \stackrel{\text{def}}{=} \mathfrak{u}$ and $\neg\mathfrak{u} \stackrel{\text{def}}{=} \mathfrak{i}$. We do not find this intuitive. Rather than being driven by a priori assumptions as to properties of negations and orders, our choice reflects our intuitions. □

DEFINITION 5. — *By an* interpretation *we mean any set of ground literals. The* truth value *of a literal $\ell$ in interpretation $\mathcal{I}$, denoted by $\mathcal{I}(\ell)$, is the value defined as follows:*

$$\mathcal{I}(\ell) \stackrel{\text{def}}{=} \begin{cases} \mathfrak{t} & \text{if } \ell \in \mathcal{I} \text{ and } (\neg\ell) \notin \mathcal{I} \\ \mathfrak{i} & \text{if } \ell \in \mathcal{I} \text{ and } (\neg\ell) \in \mathcal{I} \\ \mathfrak{u} & \text{if } \ell \notin \mathcal{I} \text{ and } (\neg\ell) \notin \mathcal{I} \\ \mathfrak{f} & \text{if } \ell \notin \mathcal{I} \text{ and } (\neg\ell) \in \mathcal{I}. \end{cases}$$

*The definition of interpretation is extended for formulas built from literals using $\vee, \wedge, \neg$ and $\to$ according to Table 1.*

Observe that in our approach we deal with direct contradictions, but also with more general notion of inconsistency. Namely, as discussed in [CM02],

> "the contradictoriness of a given theory/logic was to be identified with the fact that it derives at least some pairs of formulas of the form $A$ and $\neg A$, while inconsistency was usually talked about as a model-theoretic property to be guaranteed so that our theories can make sense and talk about 'real existing structures'[...]".

Inconsistencies may appear in our approach not only due to literals contradicting each other. Heads of rules may be assigned the value $\mathfrak{i}$ by applications of rules with inconsistent bodies. In the rest of the paper we then prefer to talk about more general concept of inconsistencies rather than limited only to contradictions.

### 3.3. ***Declarative Semantics of 4QL***

The declarative semantics of 4QL is defined in terms of Herbrand models. In more traditional approaches Herbrand models are intensively studied (see, e.g., [CGT89]). Our ap-proach is based on well-supported models. Since for any set of 4QL rules there is a unique such model (see Theorem 15), it provides truth values for all formulas. Note that interpretations we deal with may contain negative literals. Therefore we assume that *Herbrand bases* consist of all (positive and negative) ground literals.

DEFINITION 6. — *A set of literals* $\mathcal{I}$ *is a* model of a set of rules $S$, *denoted by* $\mathcal{I} \models S$, *iff for each rule* $\varrho \in S$ *we have that* $\mathcal{I}\big(body(\varrho) \rightarrow head(\varrho)\big) = \mathfrak{t}$, *where it is assumed that the empty body takes the value* $\mathfrak{t}$ *in any interpretation.*

It should be noticed that the Herbrand base is a model of any set of rules. However, our intuition is that the knowledge represented by a set of rules should be based on the explicit knowledge represented by facts. Minimal models, if exist, may not fulfill this requirement, as shown in the following example.

EXAMPLE 7. — Let $S$ be the following set of rules:

$$wait \text{ :– } overloaded \vee rest_time\,. \tag{2}$$

$$rest_time \text{ :– } wait\,. \tag{3}$$

$$\neg\, overloaded \text{ :– } rest_time\,. \tag{4}$$

$$overloaded\,. \tag{5}$$

Observe that:

– by (5), $overloaded \in \mathcal{I}$
– now, by (2), $wait \in \mathcal{I}$ thus, by (3), $rest_time \in \mathcal{I}$
– therefore, by (4), $\neg\, overloaded \in \mathcal{I}$.

Thus in every model of $S$, $overloaded$ has the value $\mathfrak{i}$.

A minimal model of $S$ is $\mathcal{I}_{min} = \{overloaded, \neg overloaded, wait, rest_time\}$ but the only fact of $S$ (i.e., $overloaded$) has in this model value $\mathfrak{i}$ so there are no facts supporting the truth of $wait$ and $rest_time$ in this model. The intuitively correct model for $S$ is

$$\mathcal{I} = \{overloaded, \neg overloaded, wait, \neg wait, rest_time, \neg rest_time\}.$$

Namely, $wait$ obtains the value $\mathfrak{i}$ to satisfy (2). Then, to satisfy (3), $rest_time$ obtains the value $\mathfrak{i}$. □

The following definitions reflect our intuitions. Note that well-supportedness closest to ours is that of [Fag94]. However, [Fag94] concerns the classical two-valued setting.

DEFINITION 8. — *Let* $\mathcal{I}$ *be an interpretation and* $\prec$ *be a strict partial order on* $\mathcal{I}$. *Given a set of rules* $S$, *we say that a model* $\mathcal{I}$ *of* $S$ supports a rule $\varrho \in S$ w.r.t. $\prec$ *provided that:*
$body(\varrho) = \emptyset$ *or there is* $\beta_j(\varrho)$ *such that* $\mathcal{I}(\beta_j(\varrho)) = \mathfrak{t}$
*and for all literals* $\imath \in \beta_j(\varrho)$ *we have that* $\imath \prec head(\varrho)$.

DEFINITION 9. — *A model* $\mathcal{I}$ *of a set of rules* $S$ *is* well-supported *provided that there exists a strict partial order* $\prec$ *on* $\mathcal{I}$ *such that for every literal* $\ell \in \mathcal{I}$*,*

– *if* $\mathcal{I}(\ell) = \mathfrak{t}$ *then* $\mathcal{I}$ *supports* $rule(\ell)$ *w.r.t.* $\prec$*.* (6)

– *if* $\mathcal{I}(\ell) = \mathfrak{i}$ *then (at least) one of the following conditions hold:*

  – $\mathcal{I}$ *supports* $rule(\ell)$ *w.r.t.* $\prec$ (7)

  – *there is a rule* $\varrho \in \{rule(\ell), rule(\neg\ell)\}$ *with* $\mathcal{I}(body(\varrho)) = \mathfrak{i}$ *for which there is* $\beta_j(\varrho)$ *with* $\mathcal{I}(\beta_j(\varrho)) = \mathfrak{i}$ *such that for all literals* $\imath \in \beta_j(\varrho)$*,* $\imath \prec head(\varrho)$*.* (8)

REMARK 10. — In the conditions (7)–(8) one could also expect a clause concerning $\mathcal{I}(\neg\ell) = \mathfrak{i}$. On the other hand, $\mathcal{I}(\ell) = \mathfrak{i}$ implies that also $\mathcal{I}(\neg\ell) = \mathfrak{i}$ so the respective condition for $\mathcal{I}(\neg\ell) = \mathfrak{i}$ is already included in Definition 9. For example, the interpretation $\{rest, \neg rest, overloaded, \neg overloaded\}$ is a well-supported model for the set of rules:

$$\neg\, rest \text{ :– } overloaded. \tag{9}$$

$$rest. \tag{10}$$

$$overloaded. \tag{11}$$

$$\neg\, overloaded. \tag{12}$$

as well as for the set consisting of the fact

$$\neg rest.$$

together with rules (10), (11), (12). □

EXAMPLE 11 (EXAMPLE 7 CONTINUED). — The minimal model

$$\mathcal{I}_{min} = \{overloaded, \neg overloaded, wait, rest_time\}$$

of the set of rules considered in Example 7 is not well-supported. Namely, $\mathcal{I}_{min}(wait) = \mathfrak{t}$. According to Definition 9, there should be an order $\prec$ such that $\mathcal{I}_{min}$ supports rule (2). By Definition 8,

$$rest_time \prec wait. \tag{13}$$

Since $\mathcal{I}_{min}(rest_time) = \mathfrak{t}$, a similar reasoning show that we should have

$$wait \prec rest_time,$$

which together with (13) cannot hold, since $\prec$ is required to be a strict partial order. □

## 4. Properties of Well-supported Models

The following lemma explains the status of heads of rules supported by a given interpretation.

LEMMA 12. — *Let $S$ be a set of rules and $\mathcal{I}$ be a model of $S$. If $\mathcal{I}$ supports a rule $\varrho \in S$ w.r.t. a strict partial order $\prec$ then $\mathcal{I}(head(\varrho)) \in \{\mathfrak{t}, \mathfrak{i}\}$.*

PROOF — Assume that $\mathcal{I}$ supports $\varrho \in S$ w.r.t. $\prec$. By Definition 8, $body(\varrho) = \emptyset$ or there is $\beta_j(\varrho)$ such that $\mathcal{I}(\beta_j(\varrho)) = \mathfrak{t}$ and for all literals $\imath \in \beta_j(\varrho)$ we have that $\imath \prec head(\varrho)$.

In the first case $body(\varrho) = \emptyset$, so $[\ell :\!-\ .] \in S$. By Definition 6, the empty body $\emptyset$ is $\mathfrak{t}$ in any interpretation, so the rule is satisfied in $\mathcal{I}$ only when $\mathcal{I}(head(\varrho)) \in \{\mathfrak{t}, \mathfrak{i}\}$.

In the second case there is $\beta_j(\varrho)$ such that $\mathcal{I}(\beta_j(\varrho)) = \mathfrak{t}$. Similarly to the previous case we have that $\varrho$ is satisfied in $\mathcal{I}$ only when $\mathcal{I}(head(\varrho)) \in \{\mathfrak{t}, \mathfrak{i}\}$. ■

The next two lemmas (13 and 14) are used in the proof of Theorem 15 claiming the uniqueness of well-supported models.

LEMMA 13. — *Let $\mathcal{M}_1$ and $\mathcal{M}_2$ be well-supported models for a set of formulas $S$. Then, for every literal $\ell$, we have that $\mathcal{M}_1(\ell) = \mathfrak{t}$ implies $\mathcal{M}_2(\ell) \in \{\mathfrak{t}, \mathfrak{i}\}$.*

PROOF — Since $\mathcal{M}_1$ is a well-supported model, there exists a strict partial order $\prec$ on $\mathcal{M}_1$ satisfying conditions of Definition 9. Thus, for every literal $\ell \in \mathcal{M}_1$, whenever $\mathcal{M}_1(\ell) = \mathfrak{t}$, there is a rule $\varrho \in S$ with $head(\varrho) = \ell$ such that $\mathcal{M}_1$ supports $\varrho$ w.r.t. $\prec$.

We prove the lemma by induction on $\prec$.

– $body(\varrho) = \emptyset$ implies that the rule $[\ell :\!-\ .]$ is in $S$. The empty body is $\mathfrak{t}$ in any interpretation, so the rule is satisfied in $\mathcal{M}_2$ only when $\mathcal{M}_2(\ell) \in \{\mathfrak{t}, \mathfrak{i}\}$, thus the implication is trivially true

– there is $\beta_j(\varrho)$ such that $\mathcal{M}_1(\beta_j(\varrho)) = \mathfrak{t}$ and for all literals $\imath$ occurring in $\beta_j(\varrho)$ we have that $\imath \prec head(\varrho)$. Since $\beta_j(\varrho)$ is a conjunction of literals, $\mathcal{M}_1(\beta_j(\varrho)) = \mathfrak{t}$ implies that for all $\imath$ in $\beta_j(\varrho)$, $\mathcal{M}_1(\imath) = \mathfrak{t}$. By inductive assumption, for all $\imath$ in $\beta_j(\varrho)$, $\mathcal{M}_2(\imath) \in \{\mathfrak{t}, \mathfrak{i}\}$, so also $\mathcal{M}_2(\beta_j(\varrho)) \in \{\mathfrak{t}, \mathfrak{i}\}$. Of course, $\mathcal{M}_2$ has to satisfy rule $\varrho$, so for its head $\ell$ we have that $\mathcal{M}_2(\ell) \in \{\mathfrak{t}, \mathfrak{i}\}$. ■

LEMMA 14. — *Let $\mathcal{M}_1$ and $\mathcal{M}_2$ be well-supported models for a set of formulas $S$. Then, for every literal $\ell$, we have that $\mathcal{M}_1(\ell) = \mathfrak{i}$ implies $\mathcal{M}_2(\ell) \in \{\mathfrak{t}, \mathfrak{i}\}$.*

PROOF — Since $\mathcal{M}_1$ is a well-supported model, there exists a strict partial order $\prec$ on $\mathcal{M}_1$ satisfying conditions of Definition 9. We proceed by induction on $\prec$.

For every literal $\ell \in \mathcal{M}_1$, if $\mathcal{M}_1(\ell) = \mathfrak{i}$ then:

– there is a rule $\varrho$ with $head(\varrho) = \ell$ such that $\mathcal{M}_1$ supports $\varrho$ w.r.t. $\prec$; or

– there is a rule $\varrho \in S$ with $head(\varrho) \in \{\ell, \neg\ell\}$ such that there is $\beta_j(\varrho)$ satisfying $\mathcal{M}_1(\beta_j(\varrho)) = \mathfrak{i}$ and for all literals $\imath$ in $\beta_j(\varrho)$, $\imath \prec head(\varrho)$.

In the first case we have that $\mathcal{M}_1$ supports $\varrho$ w.r.t. $\prec$. If $body(\varrho) = \emptyset$ then the rule $[\ell :\!-\ .]$ is in $S$, so $\mathcal{M}_2(\ell) \in \{\mathfrak{t}, \mathfrak{i}\}$. Consider the case when there is $\beta_j(\varrho)$ such that $\mathcal{M}_1(\beta_j(\varrho)) = \mathfrak{t}$ and for all literals $\imath \in \beta_j(\varrho)$ we have that $\mathcal{M}_1(\imath) = \mathfrak{t}$ and $\imath \prec head(\varrho)$. Then, by inductive assumption, we obtain that for such literals $\mathcal{M}_2(\imath) \in \{\mathfrak{t}, \mathfrak{i}\}$. In consequence, $\mathcal{M}_2(\beta_j(\varrho)) \in \{\mathfrak{t}, \mathfrak{i}\}$. Since $\ell = head(\varrho)$, we have that $\mathcal{M}_2(\ell) \in \{\mathfrak{t}, \mathfrak{i}\}$.

It remains to consider the case when there is a rule $\varrho \in S$ with $head(\varrho) \in \{\ell, \neg\ell\}$ for which there is $\beta_j(\varrho)$ satisfying $\mathcal{M}_1(\beta_j(\varrho)) = \mathfrak{i}$ and for all literals $\imath$ in $\beta_j(\varrho)$, $\imath \prec head(\varrho)$. For all such literals, $\mathcal{M}_1(\imath) \in \{\mathfrak{t}, \mathfrak{i}\}$ and at least one of them takes the value $\mathfrak{i}$. If $\mathcal{M}_1(\imath) = \mathfrak{t}$ then by Lemma 13, $\mathcal{M}_2(\imath) \in \{\mathfrak{t}, \mathfrak{i}\}$. If $\mathcal{M}_1(\imath) = \mathfrak{i}$ then by the inductive assumption $\mathcal{M}_2(\imath) \in \{\mathfrak{t}, \mathfrak{i}\}$. Thus $\mathcal{M}_2(\beta_j(\varrho)) \in \{\mathfrak{t}, \mathfrak{i}\}$, so $\mathcal{M}_2(\ell) \in \{\mathfrak{t}, \mathfrak{i}\}$. ■

THEOREM 15. — *For any set $S$ of rules there is the unique well-supported model for $S$.*

PROOF — The fact that for every set of rules there is a well-supported model follows from Theorem 28. So here it suffices to prove that, given a set of rules $S$, there may not be two different well-supported models for $S$. Suppose the contrary, i.e., there are $\mathcal{M}_1 \neq \mathcal{M}_2$ which are well-supported models of $S$. This means that there is a literal, say $\ell$, such that it is in one of the models and is not in the other. Without loss of generality we can assume that $\ell \in \mathcal{M}_1$ and $\ell \notin \mathcal{M}_2$.

Since $\ell \in \mathcal{M}_1$, we have that $\mathcal{M}_1(\ell) \in \{\mathfrak{t}, \mathfrak{i}\}$. By Lemmas 13 and 14, $\mathcal{M}_2(\ell) \in \{\mathfrak{t}, \mathfrak{i}\}$, so $\ell \in \mathcal{M}_2$ and a contradiction is reached. ■

## 5. Computing the Unique Well-Supported Model

Let us now present an algorithm for computing the unique well-supported model for a given set of rules.

DEFINITION 16. — *Let $S$ be a set of rules.*

– *By $\mathcal{L}(S)$ we denote the set of relation symbols appearing in $S$.*
– *By a* duplicate *of a relation symbol $\ell \in \mathcal{L}(S)$ we understand a fresh relation symbol, for simplicity denoted by $\ell'$.*
– *By $\mathcal{L}'(S)$ we understand the set of* duplicates *of relation symbols of $\mathcal{L}(S)$, i.e., $\mathcal{L}'(S) = \{\ell' \mid \ell \in \mathcal{L}(S)\}$.*
– *By $Pos(S)$ we understand the* DATALOG *program obtained from $S$ by replacing each negative literal $\neg\ell$ of $S$ by its duplicate $\ell'$.*

The algorithm is shown in Figure 1. The following example illustrates its execution.

EXAMPLE 17. — To illustrate the algorithm given in Figure 1, consider set of rules discussed in Example 7 together with rules:

$$good_mood :\!- rested \vee success\,. \tag{17}$$

$$\neg rested :\!- \neg rest_time\,. \tag{18}$$

$$rested\,. \tag{19}$$

$$success\,. \tag{20}$$

Phase 1 gives $\mathcal{I}_1^S = \{overloaded, \neg overloaded\}$.

Phase 2 gives the following set of rules $S'$:

– **Input:** a set of rules $S$
– **Output:** the unique well-supported model $\mathcal{I}^S$ for $S$.

1) Phase 1 (*finding basic inconsistencies*):
   a) compute the least Herbrand model $\mathcal{I}_0^S$ of $Pos(S)$
   b) let $\mathcal{I}_1^S \stackrel{\text{def}}{=} \{\ell, \neg\ell \mid \ell, \ell' \in \mathcal{I}_0^S\}$
2) Phase 2 (*finding potentially true literals*):
   a) let $S' = \{\varrho \mid \varrho \in S$ and $\mathcal{I}_1^S\big(head(\varrho)\big) \neq \mathfrak{i}\}$
   b) set $\mathcal{I}_2^S$ to be the least Herbrand model for $Pos(S')$ with literals $\ell'$ substituted by $\neg\ell$
3) Phase 3 (*reasoning with inconsistency*):
   a) define the following transformation $\Phi^S$ on interpretations:

$$\Phi^S(\mathcal{I}) \stackrel{\text{def}}{=} \mathcal{I} \cup \big\{\ell, \neg\ell \mid \text{there is a rule } [\ell :\!-\, \beta_1 \vee \ldots \vee \beta_m] \in S \tag{14}$$

$$\text{such that } \exists k \in \{1, \ldots, m\}[\mathcal{I}(\beta_k) = \mathfrak{i}\,] \tag{15}$$

$$\text{and } \neg\exists n \in \{1, \ldots, m\}[\,(\mathcal{I}_2^S - \mathcal{I})(\beta_n) = \mathfrak{t}\,]\big\}. \tag{16}$$

   The transformation $\Phi^S$ is monotonic (see Lemma 21).
   Denote by $\mathcal{I}_3^S$ the fixpoint of $\Phi^S$ obtained by iterating $\Phi^S$ on $\mathcal{I}_1^S$, i.e.,

$$\mathcal{I}_3^S = \bigcup_{i \in \omega} (\Phi^S)^i(\mathcal{I}_1^S)$$

   b) set $\mathcal{I}^S = \mathcal{I}_2^S \cup \mathcal{I}_3^S$.

**Figure 1.** *The method of computing the well-supported model for the set of rules $S$.*

$wait$ :– $overloaded \vee rest_time$ .
$rest_time$ :– $wait$ .
$good_mood$ :– $rested \vee success$ .
$\neg rested$ :– $\neg rest_time$ .
$rested$ .
$success$ .

The resulting set $\mathcal{I}_2^S$ is $\{success, rested, good_mood\}$.

Phase 3 gives the following iterations of $\Phi^S$:

$\{overloaded, \neg overloaded\}$
$\{overloaded, \neg overloaded, wait, \neg wait\}$
$\{overloaded, \neg overloaded, wait, \neg wait, rest_time, \neg rest_time\}$
$\{overloaded, \neg overloaded, wait, \neg wait, rest_time, \neg rest_time,$
$rested, \neg rested\}$ – fixpoint.

Hence $\mathcal{I}_3^S = \{overloaded, \neg overloaded, wait, \neg wait, rest_time, \neg rest_time,$
$rested, \neg rested\}$.

Finally, $\mathcal{I}^S = \{success, good_mood, overloaded, \neg overloaded, wait, \neg wait,$
$rest_time, \neg rest_time, rested, \neg rested\}$,
which is the desired result. Namely, as discussed in Example 7,

$$overloaded, \neg overloaded, wait, \neg wait, rest_time, \neg rest_time \in \mathcal{I}.$$

By (20), $success \in \mathcal{I}$ so, by (17), $good_mood \in \mathcal{I}$. Finally, to satisfy (18), both $rest, \neg rest \in \mathcal{I}$. □

In the following lemmas and theorem we assume that $S$ is an arbitrary (finite) set of rules and use notation as in the algorithm shown in Figure 1.

LEMMA 18. — *If $\ell \in \mathcal{I}_1^S$ then the value of $\ell$ is $\mathfrak{i}$ in every model of $S$.*

PROOF — Follows directly from the fact that $\mathcal{I}_0^S$ is the least Herbrand model for $Pos(S)$. ■

LEMMA 19. — *If for a well-supported model $\mathcal{M}$ of $S$ we have $\mathcal{M}(\ell) = \mathfrak{t}$ then $\ell \in \mathcal{I}_2^S$.*

PROOF — According to Definition 9, if $\mathcal{M}$ is a well-supported model then there is a strict partial order $\prec$ such that $\mathcal{M}(\ell) = \mathfrak{t}$ implies that there is a rule

$$[\ell \coloneq b_1 \vee \ldots \vee b_m] \in S$$

and $1 \leq l \leq m$ such that $b_l = b_{l1}, \ldots, b_{li_l}, \mathcal{I}(b_l) = \mathfrak{t}$ and for $1 \leq j \leq i_l$ we have that $b_{lj} \prec \ell$. This order shows a "computation" of $S'$ forcing $\ell$ to be in $\mathcal{I}_2^S$, being the least Herbrand model for $Pos(S')$ (with literals $\ell'$ substituted by $\neg\ell$). ■

LEMMA 20. — *For any literal $\ell$, we have that $\mathcal{I}_2^S(\ell) \neq \mathfrak{i}$.*

PROOF — Suppose that $\mathcal{I}_2^S(\ell) = \mathfrak{i}$. Then, by the construction of $S'$ in Phase 2(a) of the algorithm shown in Figure 1, $\mathcal{I}_1^S(\ell) = \mathfrak{i}$. In such a case, again by the construction of $S'$, the rule making $\ell$ inconsistent is removed, so no rule with $\ell$ in its head appears in $S'$. Thus $I_2^S(\ell) \neq \mathfrak{i}$ and a contradiction is reached. ■

LEMMA 21. — *The transformation $\Phi^S$ is monotonic w.r.t. set inclusion $\subseteq$, i.e.,*

$$\mathcal{I} \subseteq \mathcal{J} \textit{ implies } \Phi^S(\mathcal{I}) \subseteq \Phi^S(\mathcal{J}).$$

PROOF — Suppose $\mathcal{I} \subseteq \mathcal{J}$ and $\Phi^S(\mathcal{I}) \not\subseteq \Phi^S(\mathcal{J})$. Then there is a literal $\ell$ such that $\ell \in \Phi^S(\mathcal{I})$ and $\ell \notin \Phi^S(\mathcal{J})$. Observe that $\ell \notin \mathcal{I}$, for otherwise $\ell \in \mathcal{J}$, so also $\ell \in \Phi^S(\mathcal{J})$. Since $\ell \in \Phi^S(\mathcal{I})$ and $\ell \notin \mathcal{I}$, there is a rule $[\ell \coloneq b_1 \vee \ldots \vee b_m] \in S$ such that:

$$\exists k \in \{1, \ldots, m\}[\mathcal{I}(b_k) = \mathfrak{i}\,] \tag{21}$$

$$\neg\exists n \in \{1, \ldots, m\}[\,(\mathcal{I}_2^S - \mathcal{I})(b_n) = \mathfrak{t}\,]. \tag{22}$$

By (21) and the assumption that $\mathcal{I} \subseteq \mathcal{J}$, we have

$$\exists k \in \{1, \ldots, m\}[\mathcal{J}(b_k) = \mathfrak{i}\,]. \tag{23}$$

Suppose that $\exists n \in \{1,\dots,m\}[\,(\mathcal{I}_2^S - \mathcal{J})(b_n) = \mathfrak{t}\,]$. By the assumption $\mathcal{I} \subseteq \mathcal{J}$ we have that $\mathcal{I}_2^S - \mathcal{J} \subseteq \mathcal{I}_2^S - \mathcal{I}$, so $(\mathcal{I}_2^S - \mathcal{J})(b_n) = \mathfrak{t}$ implies that $(\mathcal{I}_2^S - \mathcal{I})(b_n) \in \{\mathfrak{t}, \mathfrak{i}\}$. However, by Lemma 20, for any literal $\ell$, we have that $\mathcal{I}_2^S(\ell) \neq \mathfrak{i}$, so $(\mathcal{I}_2^S - \mathcal{I})(b_n) \neq \mathfrak{i}$ and we conclude that $(\mathcal{I}_2^S - \mathcal{I})(b_n) = \mathfrak{t}$, which contradicts (22). Therefore,

$$\neg\exists n \in \{1,\dots,m\}[\,(\mathcal{I}_2^S - \mathcal{J})(b_n) = \mathfrak{t}\,]. \tag{24}$$

By the definition of $\Phi^S$ and properties (23), (24) we have that $\ell, \neg\ell \in \Phi^S(\mathcal{J})$, which contradicts the assumption that $\ell \notin \Phi^S(\mathcal{J})$, i.e., that $\Phi^S(\mathcal{I}) \not\subseteq \Phi^S(\mathcal{J})$. ■

By the construction of $\mathcal{I}_3^S$, we have the following obvious proposition.

PROPOSITION 22. — *For any $\ell \in \mathcal{I}_3^S$ we have $\mathcal{I}_3^S(\ell) = \mathfrak{i}$.*

LEMMA 23. — *The interpretation $\mathcal{I}^S$ of $S$ is a model for $S$.*

PROOF — Suppose $\mathcal{I}^S$ is not a model. Then there is a rule $[\ell \coloneq b_1 \vee \ldots \vee b_m] \in S$ such that either

(i) $\mathcal{I}^S(b_1 \vee \ldots \vee b_m) = \mathfrak{t}$ and $\mathcal{I}^S(\ell) \in \{\mathfrak{u}, \mathfrak{f}\}$, or
(ii) $\mathcal{I}^S(b_1 \vee \ldots \vee b_m) = \mathfrak{i}$ and $\mathcal{I}^S(\ell) \in \{\mathfrak{u}, \mathfrak{f}, \mathfrak{t}\}$.

Case (i) cannot hold since in Phase 2, the algorithm would assign $\mathfrak{t}$ to $\ell$ in $\mathcal{I}_2^S$ and $\mathcal{I}_2^S \subseteq \mathcal{I}^S(\ell)$.

Case (ii) assumes that $\mathcal{I}^S(b_1 \vee \ldots \vee b_m) = \mathfrak{i}$ so the considered rule does not participate in Phase 2 of the algorithm. It is the only rule with $\ell$ in the head. Therefore, the rule cannot be assigned $\mathfrak{t}$, so by Definition 6 and semantics of implication given in Table 1, the rule is assigned $\mathfrak{f}$.

Thus both cases, (i) and (ii), lead to contradiction. ■

In the rest of this section we prove that the interpretations obtained by our method are well-supported models.

DEFINITION 24. — *Let $S$ be a set of rules. We define a family $\{L_j^S \mid j \in \omega\}$ of disjoint subsets of the Herbrand base of $S$ as follows:*

– *$L_0^S$ is the set of facts in $S$* (25)

– *$L_{i+1}^S$ is the set of all literals $\ell$ such that $rule(\ell)$ has a body component $\beta_r(rule(\ell))$ consisting exclusively of literals in $\bigcup_{k \leq i} L_k^S$.* (26)

Note that $rule(\ell)$ considered in the above definition may have several body components $\beta_n(rule(\ell))$ satisfying (26).

To show that $\mathcal{I}$ constructed by our algorithm is well-supported we need a strict partial order on literals of a given set $S$ satisfying conditions of Definition 9. The required order, denoted by $\prec^S$, is defined below.

DEFINITION 25. — *The order $\lessdot^S$ is defined as the transitive closure of the following relation $\prec^S$ :*

$$\ell' \prec^S \ell \textit{ iff } \ell \in L^S_{i+1} \textit{ for some } i \geq 0 \textit{ and } \ell' \textit{ occurs in some } \beta_n(rule(\ell)) \quad \textit{consisting exclusively of literals in } \bigcup_{k \leq i} L^S_k. \tag{27}$$

EXAMPLE 26. — Consider the Herbrand base of the set of rules in Example 7. We get $L^S_0 = \{overloaded\}$, $L^S_1 = \{wait\}$, $L^S_2 = \{rest_time\}$, $L^S_3 = \{\neg overloaded\}$ and the defined strict order is $overloaded \lessdot^S wait \lessdot^S rest_time \lessdot^S \neg overloaded$. Note that:

– in the case of the minimal model $\{overloaded, \neg overloaded, wait, rest_time\}$ the above order does not satisfy condition (6) of Definition 9 since in the rule (2) there should be a body component which obtains the value $\mathfrak{t}$ and consisting of literals smaller w.r.t. $\lessdot^S$ than $wait$. The only candidate is $rest_time$ but $rest_time \not\lessdot^S wait$

– in the case of the intuitively correct model considered in Example 7, $\{overloaded, \neg overloaded, wait, \neg wait, rest_time, \neg rest_time\}$, the above order satisfies conditions of Definition 9 and the model is well-supported. □

PROPOSITION 27. — *The relation $\lessdot^S$ (see Definition 25) is a strict partial order.*

PROOF — As the sets $L^S_j(\mathcal{I})$ are disjoint, it follows by (27) that the relation $\prec^S$ is irreflexive and asymmetric. Consequently, its transitive closure $\lessdot^S$ is a strict partial order. ■

We can now formulate and prove the main result.

THEOREM 28. — *For any set of rules $S$, its model $\mathcal{I}^S$ constructed by algorithm given in Figure 1 is well-supported.*

PROOF — By Proposition 27, $\lessdot^S$ is a strict partial order. We shall show that it satisfies conditions of Definition 9 with $\prec$ replaced by $\lessdot^S$ .

We have two cases:

The case when $\mathcal{I}^S(\ell) = \mathfrak{t}$.

Note that $\mathcal{I}^S(\ell) = \mathfrak{t}$ only when $\ell \in \mathcal{I}^S_2 - \mathcal{I}^S_3$. Since $\mathcal{I}^S_2$ is the least Herbrand model for $Pos(S')$ with literals $\ell'$ substituted by $\neg\ell$, the rule $rule(\ell)$ has a body component evaluated to $\mathfrak{t}$, with all literals smaller w.r.t. $\lessdot^S$ than $\ell$, which has not been changed by iterating $\Phi$ in Phase 3. This shows condition (6) of Definition 9.

The case when $\mathcal{I}^S(\ell) = \mathfrak{i}$.

If $\mathcal{I}^S(\ell) = \mathfrak{i}$ then $\ell \in \mathcal{I}^S_1$ or $\ell \in \mathcal{I}^S_3 - \mathcal{I}^S_1$.

Consider first the case when $\ell \in \mathcal{I}^S_1$. The interpretation of $\ell \in \mathcal{I}^S_1$ is obtained from the least Herbrand model of $Pos(S)$. Therefore $\ell$ and $\neg\ell$:

(a) both satisfy condition (7) (when they both have been derived using rules with body components evaluated to $\mathfrak{t}$); or

(b) both satisfy condition (8) (when they both have been derived using rules with body components evaluated to $\mathfrak{i}$); or

(c) one of them satisfies condition (7) and the other satisfies condition (8).

In the case (a) we have that $\ell$ satisfies (7). In cases (b) and (c), $\ell$ satisfies (8).

It remains to consider the case when $\ell \in \mathcal{I}_3^S - \mathcal{I}_1^S$, i.e., when $\ell$ has been added to $\mathcal{I}_3^S$ by iterating $\Phi$ in Phase 3. This happens when the body of $rule(\ell)$ (or $rule(\neg\ell)$) became evaluated to $\mathfrak{i}$. There are two cases:

– $\ell \in \mathcal{I}_2^S$: a body component $\beta_i(rule(\ell))$ (or $\beta_i(rule(\neg\ell))$) has been $\mathfrak{t}$ and all literals in that component have been smaller w.r.t. $\prec^S$ than $\ell$. This $\beta_i(rule(\ell))$ (or $\beta_i(rule(\neg\ell))$) satisfies the condition (7) of Definition 9; or

– $\ell \notin \mathcal{I}_2^S$: a body component $\beta_i(rule(\ell))$ (or $\beta_i(rule(\neg\ell))$) has been $\mathfrak{i}$ and all literals in that component have been smaller w.r.t. $\prec^S$ than $\ell$. This $\beta_i(rule(\ell))$ (or $\beta_i(rule(\neg\ell))$) satisfies the condition (8) of Definition 9. ■

REMARK 29. — Observe that Theorem 28 provides a PTIME method for verifying whether a given model is well-supported since $\prec^S$ can be constructed in deterministic polynomial time. Definition 9 does not provide a direct PTIME method due to existential quantification over $\prec^S$. □

## 6. Layered Architecture

In this section we introduce and discuss external literals allowing us to express non-monotonic rules. The idea is similar to stratification but the problem is not with negation but with external literals. In this section we use well-known techniques related to stratified DATALOG$^\neg$. All necessary definitions and theorems related to stratifications can be found, e.g., in [AHV96]. External literals have also been considered in [VMS09].

The architecture discussed in this section is equivalent to the modular architecture of [MS11]. However, it simplifies the proof of Lemma 37 and also provides intuitions related to well-known stratified programs.

Let $\mathbb{M}, \mathbb{R}$ be disjoint sets. In what follows we assume that all relation symbols are of the form $M.R$, where $M \in \mathbb{M}$ and $R \in \mathbb{R}$. Intuitively, members of $\mathbb{M}$ are names of "modules" or "services" and members of $\mathbb{R}$ are classical relation symbols.

DEFINITION 30. — *An* external literal *is an expression of one of the forms:*

$$M.R,\ \neg M.R,\ M.R \text{ IN } T,\ \neg M.R \text{ IN } T,$$

*where:*

– $M \in \mathbb{M}$ *is called the* reference module *of the external literal;*
– $T \subseteq \{\mathfrak{t}, \mathfrak{f}, \mathfrak{i}, \mathfrak{u}\}$ *(if $T = \emptyset$ then $\ell$ IN $T$ is $\mathfrak{f}$).*

*We write $\ell = v$ to stand for $\ell$ IN $\{v\}$. For an extended literal $\ell$, by $module(\ell)$ we denote its reference module.*

The literal $\neg M.R$ IN $T$ is to be read as "$(\neg M.R)$ IN $T$" rather than "$\neg(M.R \text{ IN } T)$".

DEFINITION 31. — *By an* extended rule *we understand a rule $\varrho$ of the form* (1)*, where relation symbols are replaced by external literals assuming that $head(\varrho)$ is of the form $M.R$ or $\neg M.R$.*

The following example shows a possible use of external literals.

EXAMPLE 32. — The following rules locally close $loc$, where $loc(X, Y, T)$ means that object $X$ has location $Y$ at timepoint $T$:

| | | |
|---|---|---|
| $K.loc(X, Y, T)$ :– | $L.nextTime(T, S)$, | – $T$ is the timepoint next to $S$ |
| | $L.house(X)$, | – $X$ is a house |
| | $L.loc(X, Y, S)$, | – location of $X$ at time $S$ is $Y$ |
| | $L.chLoc(X, S)$ IN $\{\mathfrak{u}, \mathfrak{f}\}$. | – location change of $X$ is $\mathfrak{u}$ or $\mathfrak{f}$. |

Intuitively, the above rule states that houses do not change their location no matter whether $M$'s database contains information as to the change of location or not. □

DEFINITION 33. — *A set of rules* $S$ *is* well-layered *iff there is a mapping* $\kappa_S : \mathbb{M} \longrightarrow \omega$ *such that for every rule* $\varrho \in S$,

– *if* $\ell$ *is a literal of the form* $M.R$ *or* $\neg M.R$ *appearing in* $body(\varrho)$,

$$\kappa_S(module(head(\varrho))) \geq \kappa_S(module(\ell))$$

– *if* $\ell$ *is a literal of the form* $M.R$ IN $T$, $\neg M.R$ IN $T$ *appearing in* $body(\varrho)$,

$$\kappa_S(module(head(\varrho))) > \kappa_S(module(\ell)).$$

By an immediate adaptation of the algorithm for checking stratifiability of a set of rules of DATALOG$^\neg$ (see, e.g., [AHV96]), we have the following proposition.

PROPOSITION 34. — *Checking whether a set of rules* $S$ *is well-layered takes time polynomial in the size of* $S$.

We now have the following definition.

DEFINITION 35. — *An* extended 4QL program *is any finite well-layered set of extended rules.*

## 7. Complexity Issues

### 7.1. *Expressing Stratified Datalog*$^\neg$

Let $P$ be a stratified DATALOG$^\neg$ program. Let $\{1, 2, \ldots, n\}$ be all strata of $P$. First, we replace all rules with the same head $\ell$ as a single rule whose body is the disjunction of all bodies of rules with head $\ell$. This replacement is correct, since here we deal with the standard two-valued semantics for stratified DATALOG$^\neg$ programs.

For each stratum $i = 1, \ldots, n$ of $P$ we take distinct symbols $M_i, N_i \in \mathbb{M}$ and replace the stratum by:

– rules obtained from rules appearing in $i$-th stratum of $P$ by replacing each relation symbol $R$ by:
  - $M_i.R$ if $R$ is defined in stratum $i$
  - $N_j.R$ if $R$ is defined in stratum $j$ with $j < i$

– for each relation $R$ defined in stratum $i$ we add the following rules, closing the relation $R$ as the Closed World Assumption CWA does:

$$N_i.R :\!- M_i.R = \mathfrak{t}.$$
$$\neg N_i.R :\!- M_i.R \text{ IN } \{\mathfrak{f}, \mathfrak{u}\}.$$

The result of $p$ appearing in $i$-th stratum of $P$ is given by $N_i(p)$.

Observe that the resulting set of rules is an extended 4QL program which can never lead to inconsistency.

EXAMPLE 36. — Consider a stratified program, consisting of three strata, shown in the first column of Table 2. Using the method described above we obtain the equivalent 4QL program shown in the second column of Table 2, where:

– $N_1.r$ provides the value of $r$
– $N_2.s$, $N_2.q$ provide values respectively for $s$ and $q$
– $N_3.p$ provides the value of $p$. □

**Table 2.** *A stratified program and its 4QL representation.*

|  | |  | |
|---|---|---|---|
| | | $N_3.p :\!- M_3.p = \mathfrak{t}.$ | – CWA$(p)$ |
| | | $\neg N_3.p :\!- M_3.p \text{ IN } \{\mathfrak{f}, \mathfrak{u}\}.$ | – CWA$(p)$ |
| | | $M_3.p :\!- \neg N_2.q \vee N_1.r.$ | – layer 3 |
| | | $N_2.q :\!- M_2.q = \mathfrak{t}.$ | – CWA$(q)$ |
| $p :\!- \neg q.$ | – stratum 3 | $\neg N_2.q :\!- M_2.q \text{ IN } \{\mathfrak{f}, \mathfrak{u}\}.$ | – CWA$(q)$ |
| $p :\!- r.$ | – stratum 3 | $N_2.s :\!- M_2.s = \mathfrak{t}.$ | – CWA$(s)$ |
| $q :\!- r.$ | – stratum 2 | $\neg N_2.s :\!- M_2.p \text{ IN } \{\mathfrak{f}, \mathfrak{u}\}.$ | – CWA$(s)$ |
| $s :\!- q.$ | – stratum 2 | $M_2.q :\!- N_1.r.$ | – layer 2 |
| $r.$ | – stratum 1 | $M_2.s :\!- M_2.q.$ | – layer 2 |
| | | $N_1.r :\!- M_1.r = \mathfrak{t}.$ | – CWA$(r)$ |
| | | $\neg N_1.r :\!- M_1.r \text{ IN } \{\mathfrak{f}, \mathfrak{u}\}.$ | – CWA$(r)$ |
| | | $M_1.r.$ | – layer 1 |

Due to the above construction we have the following lemma.

LEMMA 37. — *Every stratified* DATALOG$^{\neg}$ *program can be expressed by an extended 4QL program.*

### 7.2. *Complexity of Layered 4QL*

First observe that Algorithm provided in Figure 1 involves standard DATALOG computations in Phase 1 and Phase 2 and a fixpoint computation in Phase 3. Such computations have PTIME complexity (see, e.g., [AHV96]). Moreover, extended 4QL programs can be constructed layer by layer, starting from the lowest layer. Therefore we have the following theorem.

THEOREM 38. — *4QL with modules has* PTIME *data complexity.*

Since stratified DATALOG captures PTIME on ordered structures (see, e.g., Theorem 15.4.8 in [AHV96]), by Lemma 37 we have that extended 4QL captures PTIME queries.

THEOREM 39. — *Extended 4QL captures* PTIME *queries on ordered structures.*

## 8. Conclusions

In the paper we have investigated the query language 4QL originally outlined in [MS11]. We defined a semantics based on the notion of well-supported model and we proposed an algorithm for computing well-supported models. The language is simple yet powerful.

We focussed on logical foundations and complexity of 4QL, proving the correctness of the algorithm for computing well-supported models and showing that 4QL has PTIME data complexity and captures PTIME.

In summary,

– 4QL provides a very flexible mechanism for dealing with the lack of knowledge and resolving possible inconsistencies in an application dependent manner
– 4QL is powerful enough to express large classes of nonmonotonic rules known from the literature (see [MS11])
– 4QL can be used as a rule language for Semantic Web and robotics applications.

There are still interesting questions concerning 4QL. Apart from implementing and applying 4QL, there are still many theoretical issues. Perhaps the most important are:

– provide a top-down method for query answering running in deterministic polynomial time
– extend the method to the case of infinite domains
– provide optimization techniques improving the performance of query answering, not necessarily by computing the whole well-supported model
– provide techniques for computing the well-supported model after a database update, on the basis of the model computed before the update.

## Acknowledgments

We would like to thank Wlodek Drabent and two anonymous reviewers for critical reading of this paper and supplying many helpful comments and remarks.

This work has been supported in part by the Polish MNiSW grant N N206 399334.